\documentstyle[12pt]{article}
\begin{document}

\noindent Stockholm\\
August 2008\\
Revised October 2008\\

\vspace{4mm}

\begin{center}

{\Large THE FRAME POTENTIAL, ON AVERAGE}

\vspace{10mm}

{\large Ingemar Bengtsson}\footnote{Email address: ingemar@physto.se. 
Supported by VR.}

\

{\sl Stockholm University, AlbaNova\\
Fysikum\\
S-106 91 Stockholm, Sweden}

\vspace{8mm}

{\large Helena Granstr\"om}\footnote{Email address: helenag@math.su.se.}

\

{\sl Stockholm University\\
Institutionen f\"or matematik\\
S-106 91 Stockholm, Sweden}

\vspace{10mm}

{\bf Abstract}

\end{center}

\vspace{5mm}

\noindent A SIC consists of $N^2$ equiangular unit vectors in an $N$ dimensional 
Hilbert space. The frame potential is a function of $N^2$ unit vectors. It has a unique 
global minimum if the vectors form a SIC, and this property has been made use of in numerical 
searches for SICs. When the vectors form an orbit of the Heisenberg group 
the frame potential becomes a function of a single fiducial vector. We analytically compute the 
average of this function over Hilbert space. We also compute averages when the 
fiducial vector is placed in certain special subspaces defined by the Clifford 
group.  

\newpage

{\bf 1. Introduction}

\vspace{5mm}

\noindent Symmetric Informationally Complete Positive Operator Valued Measures, 
or SICs, is the unwieldy name for a simple idea \cite{Zauner, Renes}: a set 
of $N^2$ unit vectors 
in an $N$ dimensional Hilbert space, equiangular in the sense that 

\begin{equation} |\langle \psi_I|\psi_J\rangle |^2 = \frac{1}{N+1} \ , 
\hspace{8mm} 1 \leq I, J \leq N \ , \hspace{4mm} I \neq J \ . \end{equation}

\noindent If the vectors are reinterpreted as projectors, that is as points 
in the set of Hermitian matrices of unit trace (the space where the density 
matrices live), they form a regular simplex in an $N^2-1$ dimensional Euclidean space. 
This also explains why we want their number to be $N^2$. The corners of such 
a simplex can be used to define barycentric coordinates for any density matrix, 
which is what ``informationally complete'' stands for. In quantum information 
theory SICs have attracted attention because they---if they exist---are 
useful for quantum state tomography \cite{Renes, Scott}. In quantum foundations 
they have attracted 
attention as a preferred kind of measurement \cite{Chris}---and they have been 
studied in many other branches of science under names such as ``equiangular 
lines'' \cite{equi}, ``equiangular tight frames'' \cite{Benedetto}, and ``maximal 
quantum designs'' \cite{Zauner}. Strohmer and Heath provide a mathematical 
survey \cite{Heath}.    

The question whether SICs exist for all $N$ has turned out to be extraordinarily 
difficult to answer. They have been constructed in most (but not all) dimensions $N \leq 19$ 
\cite{Zauner, Grassl, Marcus, Markus, MG}. Numerical searches have been successful 
for all $N \leq 45$ 
\cite{Renes}, but no general formula has emerged. This is a bit surprising, 
given that we are really asking a very simple question about the shape of the 
convex body of density matrices: is it possible to inscribe a regular simplex 
in this body, with $N^2$ corners on its outsphere?  
The available evidence does however suggest that the answer is ``yes'' for all $N$, 
and moreover one can always find a SIC covariant under the 
Heisenberg-Weyl group, meaning that it can be constructed by first choosing 
a fiducial vector $|\psi_0\rangle$, and then acting on this vector with the 
$N^2$ elements of the (projective) Heisenberg-Weyl group. 

A natural measure of how close a given set of $N^2$ unit vectors $|\psi_I\rangle $ 
is to forming a SIC is given by the function 

\begin{equation} f \equiv \frac{1}{2}\sum_{I \neq J}\left( | \langle \psi_I|\psi_J\rangle 
|^2 - \frac{1}{N+1}\right)^2 \ . \label{f} \end{equation}
 
\noindent By construction $f = 0$ if and only if the vectors form a SIC. The function 
$f$ is related to the frame potentials 

\begin{equation} F_t \equiv \sum_{I,J}|\langle \psi_I|\psi_J\rangle |^{2t} 
\ , \hspace{8mm} t \in \{ 1,2\}  
\end{equation}

\noindent by 

\begin{equation} f = \frac{1}{2}F_2 - \frac{1}{N+1}F_1 \ . \end{equation}

\noindent These frame potentials also assume their global minimum for a SIC, and in fact 
it is enough if $F_2$ assumes its minimum, since $F_1$ is known to follow suit. 
Therefore numerical searches for SICs have focused on minimizing $F_2$. Actually a 
frame potential is defined for any integer $t$ \cite{Renes, Benedetto}, but this does 
not concern us here. 

We will have nothing to say about the existence problem here, rather we will 
compute averages of the function $f$, with and without the assumption of group 
covariance. We use the Fubini-Study measure on complex projective space to 
perform the averaging. We also compute averages when the fiducial vector is 
restricted to lie in certain subspaces defined by 
the Heisenberg group and its normalizer (the so-called Clifford group). 
These calculations were made for a reason: in the course of an investigation 
of certain configurations of $N^2$ vectors that occur in a related problem 
(to be precise but perhaps not informative, the torsion points of an elliptic 
curve defined by a complete set of mutually unbiased bases) we computed 
the values of $f$ for these configurations. And the question then arose whether 
the values we obtained were high or low. To answer questions like this---and 
there are many such questions---one needs to know these averages. 

Our final results have a certain elegance. Our paper is organized as follows: 
in section 2 we state some definitions, and in section 3 we describe the special 
subspaces we are interested in. In section 4 we give details concerning our 
calculations, which were carried out by brute force. Readers who are familiar 
with the SIC problem, and 
who do not want the details of an involved calculation, are advised to go 
directly to section 5, where we state our results and make some comments.\footnote{The first 
version of this paper contained a mistake, in eq. (\ref{fel}). A counterintuitive and 
exciting result ensued. The mistake was spotted by Christopher 
Fuchs and by an anonymous referee; it is better to be right than to be exciting, and 
we thank them for this.}
 
\vspace{1cm}

{\bf 2. Definitions}

\vspace{5mm}

\noindent The Heisenberg-Weyl group \cite{Weyl} is the group generated by two 
elements $\tau$ and $\sigma$ subject to the relations 

\begin{equation} \sigma \tau = q\tau \sigma  \ , \hspace{5mm} \tau^N = \sigma^N 
= 1 \ , \hspace{8mm} q \equiv e^{2\pi i/N} \ . \end{equation}

\noindent Up to unitary equivalence there is a unique unitary representation of this 
group such that 

\begin{equation} \sigma |a\rangle = q^a |a\rangle \hspace{12mm} \tau |a\rangle = | a+1 \rangle 
\ , \hspace{8mm} 
0 \leq a \leq N-1 \ . \end{equation}

\noindent This is how we fix coordinates in Hilbert space; vectors can then be defined 
by their components

\begin{equation} |\psi\rangle = \sum_{a=0}^{N-1}Z^a|a\rangle \ . \end{equation}

\noindent Up to phases, a general group element is 

\begin{equation} D_{ij} = \tau^i\sigma^j \ , \hspace{8mm} 0 \leq i,j \leq N-1 \ . \end{equation} 

\noindent The phases do not matter to us. 

%
%

Now consider $N^2$ unit vectors forming an orbit under the Heisenberg group, 

\begin{equation} |\psi_{ij}\rangle = D_{ij}|\psi_0\rangle \ , \end{equation}

\noindent where $|\psi_0\rangle $ is some fiducial unit vector. The frame 
potential evaluated on such an orbit becomes a function on the projective 
Hilbert space ${\bf CP}^{N-1}$. This is the case we are most interested in, 
so we define 

\begin{equation} f_H = \frac{N^2}{2}\sum_{i,j \neq 0,0}\left( 
| \langle \psi_0|\psi_{ij}\rangle |^2 - \frac{1}{N+1}\right)^2 \ . 
\label{fH} \end{equation}

\noindent Because of group covariance there are only $N^2-1$ terms in the sum.  

The definition of $f_H$ can be manipulated further \cite{Chris, Mahdad}. Following 
Appleby, Dang and Fuchs we use an explicit matrix representation of $D_{ij}$ to write 

\begin{equation} \langle \psi_0|D_{ij}|\psi_0\rangle = \bar{Z}_aq^{bj}\delta_{b+i}^a 
Z^b = q^{(a-b)j}\bar{Z}_aZ_{a-i} \ . \end{equation}

\noindent Then we take the Fourier transform

\begin{equation} \frac{1}{N}\sum_jq^{kj}|\langle \psi_0|D_{ij}|\psi_0\rangle|^2 = 
\sum_a\bar{Z}_a\bar{Z}_{a+k-i}Z_{a+k}Z_{a-i} \ . \end{equation}

%
%
\noindent From this it is easy to show that the first frame potential $F_1 = N^3$ 
(this is true for all SIC-POVMs), and moreover that 

\begin{equation} f_H = \frac{N^3}{2}\left( \sum_{i=0}^{N-1}\sum_{k=0}^{N-1}
\left| \sum_{a=0}^{N-1}\bar{Z}_a\bar{Z}_{a+k-i}Z_{a+k}Z_{a-i}\right|^2 - 
\frac{2}{N+1}\right) \ . \label{sum} \end{equation}

\noindent This is the expression we will work with in the sequel.  

We will be interested in averages of $f$ and $f_H$. To compute these averages 
we use the Fubini-Study measure ${\rm d}\mu_{\rm FS}$ on projective Hilbert 
space \cite{BZ}; this is the natural definition of an average in the absence 
of any special information. What we wish to compute is   

\begin{equation} \left< f\right> \equiv \frac{1}{\mbox{vol}[{\bf CP}^{N-1}]}
\int {\rm d}\mu_{\rm FS} \ f \ , \end{equation}

\noindent and similarly for $f_H$. Computing $\left< f\right>$ is straightforward: 
using first the linearity of the expectation value and then the unitary 
invariance of the measure we obtain 

\begin{equation} \left< f\right> = \frac{N^2(N^2-1)}{2}\left< \left( |\langle \psi_0|
\psi\rangle |^2 - \frac{1}{N+1}\right)^2 \right> \ . \end{equation}

\noindent Here $|\psi_0\rangle$ is any fixed vector. To compute 
$\left< f_H\right>$ requires more work---we will fall back on 
the explicit expression (\ref{sum}), and then collect terms. 

We have now defined the functions we wish to average, and we have defined 
``average''. It remains to define the special subspaces of Hilbert space 
that we are about to consider. 

\vspace{1cm}

{\bf 3. Special subspaces}

\vspace{5mm}

\noindent We will compute averages of $f_H$ also when the fiducial vector is confined 
to lie in certain interesting subspaces of the Hilbert space, picked out by the 
Clifford group. By definition the latter is the normalizer of the Heisenberg-Weyl group 
in the group of all unitaries. Thus the Clifford group is the group of all unitaries 
$U$ such that 

\begin{equation} UD_{ij}U^\dagger = \omega D_{i'j'} \ , \end{equation}

\noindent where $\omega$ is a phase factor. We are interested in representations up 
to a phase, and it can be shown that the Clifford group with the phases ignored is 
isomorphic to a semi-direct product of a symplectic group with the Heisenberg group 
itself. The importance of this automorphism group was stressed by Zauner \cite{Zauner} 
and Grassl \cite{Grassl}; for a self-contained account see Appleby \cite{Marcus}. 

When the dimension is odd the Clifford group contains elements of order 2. They play 
a major role in the definition of discrete Wigner functions for the odd dimensional 
case \cite{Subhash, Gross}, and for the special case when the dimension is an odd 
prime number they are also known as Wootters' phase point operators \cite{Wootters}.
They are symmetries of an elliptic curve associated to a complete set 
of mutually unbiased bases \cite{Hulek}. An elliptic curve can be defined as an 
embedding of a torus into complex projective space. The Heisenberg group acts on 
this torus, and gives rise to $N^2$ torsion points on the curve, roughly analogous 
to the $N$th roots of unity on a circle. Hughston \cite{Hughston} has made the 
interesting comment that, in the special case $N = 3$, these torsion points define 
a SIC. In general it is known that the torsion points lie in the $N^2$ subspaces 
defined by elements of the Clifford group of order 2. In particular there is such 
an element $A$ obeying 

\begin{equation} AD_{ij}A = D_{-i,-j} \ . \end{equation}

\noindent Because $A^2 = 1$, the operator $A$ is both unitary and Hermitean. 
It splits Hilbert space into two subspaces ${\cal H}^\pm$ of dimensions 
$n$ and $n-1$, respectively, where $N = 2n-1$. Explicitly these subspaces are 
defined by 

\begin{equation} {\cal H}^+ : \hspace{2mm} 
\frac{1}{\sqrt{2}}\left( \begin{array}{c} \sqrt{2}x_0 \\ 
x_1 \\ \vdots \\ x_{n-1} \\ x_{n-1} \\ \vdots \\ x_1 \end{array}\right) \hspace{10mm} 
{\cal H}^- : \hspace{2mm} 
\frac{1}{\sqrt{2}}\left( \begin{array}{c} 0 \\ 
x_1 \\ \vdots \\ x_{n-1} \\ - x_{n-1} \\ \vdots \\ - x_1 \end{array}\right)
\ . \end{equation} 

\noindent Unfortunately the fiducial vectors of the SICs do not lie in these 
subspaces when $N > 3$. However, we believe that $f_H$ averaged using the Fubini-Study 
measure in such subspaces gives some feeling for the systematics of the SIC problem, 
and we will compute this average.

For any $N$ the Clifford group contains elements of order 3. Zauner \cite{Zauner} 
conjectured that there always exists a SIC such that the fiducial vector 
is an eigenvector of a symplectic transformation of order 3, and his conjecture 
has been verified by Appleby \cite{Marcus} in all available cases \cite{Renes}. 
We would therefore like to know the average of $f_H$ over such subspaces. Unfortunately 
it is not so easy to describe these subspaces for arbitrary $N$, and we therefore 
confined ourselves to the special case $N = 7$. Then there exists an element of order 
3 acting like a permutation matrix, and the subspaces are explicitly 

\begin{equation} {\cal H}_{1} : \hspace{2mm} 
\frac{1}{\sqrt{3}}\left( \begin{array}{c} \sqrt{3}x_0 \\ 
x_1 \\ x_1 \\ x_2 \\ x_1 \\ x_2 \\ x_2 \end{array}\right) \hspace{10mm} 
{\cal H}_{\alpha } : \hspace{2mm} 
\frac{1}{\sqrt{3}}\left( \begin{array}{c} 0 \\ 
x_1 \\ \alpha^2x_1 \\ x_2 \\ \alpha x_1 \\ \alpha x_2 \\ \alpha^2 x_2 \end{array}\right)
\end{equation}

\noindent where $\alpha$, the eigenvalue, is a primitive third root of unity. The dimension 
of the subspace ${\cal H}_{1}$ is 3, and the dimensions of the two orthogonal subspaces 
are 2. There is a fiducial vector for a SIC in ${\cal H}_{1}$ \cite{Marcus}. It seemed 
reasonable to expect that the average $f_H$ over this subspace would be quite low, but 
this expectation was not borne out.  

\vspace{1cm}

{\bf 4. Calculations}

\vspace{5mm}

\noindent In order to average the frame potential (\ref{sum}) over the entire Hilbert space, 
we observe---see eq. (\ref{integr}) below---that the angular integrals will make all non-real 
terms of the sum disappear. We 
also note that a term of the form $|Z_1|^4 |Z_2|^4$ gives the exact same contribution to the 
average as one of, say, the form $|Z_3|^4 |Z_5|^4$, and all one needs to keep track of is 
the number of different $|Z_i|$ in a term, as well as the exponents present. Consequently, 
we look at the sum (\ref{sum}), and ask in how many ways we can get a term of the type 
$|Z_1|^8$, in how many ways one of the type $|Z_1|^6|Z_2|^2$, and so on. The problem 
has then been reduced to elementary equation solving.

Computing the average of $f_H$ as given in (\ref{sum}), amounts to computing the average

\begin{equation} \label{sigma} \left< \Sigma \right> = \left< \sum_{i=0}^{N-1}\sum_{k=0}^{N-1}
\left| \sum_{a=0}^{N-1}\bar{Z}_a\bar{Z}_{a+k-i}Z_{a+k}Z_{a-i}\right|^2 \right>  \ . \end{equation}

\noindent For calculational convenience, the sum can be split into two parts: the $i=0$ part 
where all terms are real and we have to keep track of cross terms, and the remaining $i\neq 0$ 
part, where cross terms contribute only in special cases: 

\begin{equation} \label{framesum} \left< \Sigma  \right> = \sum_{k=0}^{N-1}
\left( \sum_{a=0}^{N-1}|{Z}_a|^2|{Z}_{a+k}|^2\right)^2 + \sum_{i=1}^{N-1}\sum_{k=0}^{N-1}
\left| \sum_{a=0}^{N-1}\bar{Z}_a\bar{Z}_{a+k-i}Z_{a+k}Z_{a-i}\right|^2 \ .\end{equation} 

Let us first consider the $i=0$ term, which can be rewritten as 
$$\sum_{k=0}^{N-1}\sum_{a=0}^{N-1}\sum_{b=0}^{N-1}|{Z}_a|^2|{Z}_{a+k}|^2 
|{Z}_b|^2|{Z}_{b+k}|^2 .$$
We will divide our analysis into two cases, the case $a=b$ and the case $a \neq b$. 
When $a=b$ the terms we get will be either of the type $|Z_1|^8$---this will occur for 
$k=0$---or of the type $|Z_1|^4 |Z_2|^4$, which will be the case for all remaining values 
of $k$, no matter what is the value of $a$ and $b$. The total contribution, modulo integration, 
from the $i=0$ term will for $a=b$ then be
\begin{equation} N |Z_1|^8 + N(N-1) |Z_1|^4 |Z_2|^4, \end{equation}
where the factor $N$ in front of the first term comes from the number of possible choices 
of $a=b$, and 
the second term is multiplied by the factor $N(N-1)$ for $N$ different choices of $a=b$ and $N-1$ 
choices of $k \neq 0$.

We now turn to the sum for different values of $a$ and $b$. In this case, we will have four 
different types of $|Z_i|$, each to the power of two, in each term, except in the three special 
cases where the values of two or more of the indices coincide, namely when $a=b+k$, $b=a+k$ and 
$k=0$. In the first two cases we get terms of the $|Z_1|^4|Z_2|^2|Z_3|^2$ type, in the last we 
get a $|Z_1|^4 |Z_2|^4$ type term. The remaining choices of $k$, $a$ and $b$---there are 
$N(N-1)(N-3)$ of them---give terms with all indices different, i.e. terms of the type 
$|Z_1|^2 |Z_2|^2|Z_3|^2|Z_4|^2$. The total contribution from the $i=0$ term is given in 
table \ref{tabell}.

Moving on to the $i \neq 0$ case, we can forget about cross terms except when $k=0$. 
Consequently, when $k \neq 0$ we can consider the sum
$$ 
\sum_{i=1}^{N-1}\sum_{k=1}^{N-1} \sum_{a=0}^{N-1} |{Z}_a|^2 |Z_{a+k-i}|^2 |Z_{a+k}|^2 |Z_{a-i}|^2,
$$
while when $k=0$ we still have to work with the expression from equation (\ref{framesum}). 
As in the previous case, we ask what special cases occur, i.e. when two or more indices take 
on equal values. These cases turn out to be $k=i$, $k=-i$ and $k=0$. The respective 
contributions to the average are given in the table, as is the contribution for all other 
choices of indices. 

\begin{table}[!h]
\begin{center}
\begin{tabular}{|l|c|c|c|c|c|}
\hline
 & $Z_1^8$ & $Z_1^6Z_2^2$ & $Z_1^4Z_2^4$ & $Z_1^4Z_2^2Z_3^2$ &
$Z_1^2Z_2^2Z_3^2Z_4^2$ \\ \hline

$i=0, a = b$ & $N$ & --- & $N(N-1)$  & --- & --- \\ \hline
$i=0, a \neq b, k=0$ & --- & --- & $N(N-1)$ &---  &---  \\ \hline
$i=0, a \neq b, k=a-b$ & --- & --- & --- & $N(N-1)$ & --- \\ \hline
$i=0, a \neq b, k=b-a$ & --- & --- & --- & $N(N-1)$ & --- \\ \hline
$i=0, \textrm{others} $ &---  & --- &---  & --- & $N(N-1)(N-3)$ \\ \hline

$i \neq 0, k=0$ & --- & --- & $N(N-1)$ & $2N(N-1)$ & $N(N-1)(N-3)$ \\
\hline 
$i \neq 0, k=i$ & --- & --- & --- & $N(N-1)$ & --- \\
\hline 
$i \neq 0, k= - i$ & --- & --- & --- & $N(N-1)$ & --- \\
\hline 
$i \neq 0, \textrm{others}$ & --- & --- & --- & --- & $N(N-1)(N-3)$ \\
\hline 
total & $N$ & --- & $3N(N-1)$ & $6N(N-1)$ & $3N(N-1)(N-3)$ \\
\hline 
\end{tabular}
\caption{Number of different type terms in the average for odd $N$.}
\label{tabell}
\end{center}
\end{table}

Subtracting the constant term and multiplying by the initial factor of (\ref{sum}) yields 
an expression which can be integrated over all of Hilbert space in order to obtain the 
SIC-function average. 
To perform the integrals we parametrize the unit vectors as 

\begin{equation} Z^a = (\sqrt{p_0}, \sqrt{p_k}e^{i\nu_k}) \ , \end{equation}

\noindent where the ranges of the parameters are 

\begin{equation} p_0 + p_1 + \dots + p_{N-1} = 1 \ , \ \ p_0 \geq 0 \ , \ \ p_k 
\geq 0 \ , \hspace{8mm} 0 \leq \nu_k 
< 2\pi \ . \end{equation} 

\noindent We solve for $p_0$, and obtain the explicit expression \cite{BZ} 

\begin{equation} \left< f \right>  = 
\frac{(N-1)!}{(2\pi)^{N-1}}\int_0^1 {\rm d}p_1 \dots \int_0^{1-p_1 - \dots - p_{N-2}}
{\rm dp}_{N-1}
\int_0^{2\pi}{\rm d}\nu_1 \dots \int_0^{2\pi}{\rm d}\nu_{N-1} \ f \ . \label{integr} \end{equation}

In calculating the averages a number of standard integrals will be used, namely\\

\begin{equation} \left< |Z_1Z_2Z_3Z_4|^2\right> =  \frac{(N-1)!}{(N+3)!}  \label{2-2-2-2} \end{equation}
 
\begin{equation} \left< |Z_1|^4|Z_2Z_3|^2\right> = 2\frac{(N-1)!}{(N+3)!}
\end{equation}

\begin{equation} \left< |Z_1Z_2|^4\right> = 2^2\frac{(N-1)!}{(N+3)!} \end{equation}

\begin{equation} \left< |Z_1|^6|Z_2|^2\right> = 3!\frac{(N-1)!}{(N+3)!} \label{6-2} 
\end{equation}

\begin{equation} \left< |Z_1|^8\right> = 4!\frac{(N-1)!}{(N+3)!} \ . \label{8} \end{equation}

The computations for odd values of $N$ are straightforward, with few complications. Even $N$ are, 
however, a somewhat different story. To begin with, cross terms play in also for non-zero values of 
$i$ and $k$, and in a somewhat different manner than in the odd $N$ case. Whereas we get cross term 
contributions for odd $N$ only when all terms are real in a sum---as for $k=0$, for instance---we 
here find pairwise equal terms inside the modulus sign of the $i \neq 0$ term of equation 
(\ref{framesum}), so that care has to be exercised when squaring. Using the same strategy as 
for odd $N$---dividing the sum into $i=0$ and $i \neq 0$ parts and asking when two or more 
indices are equal---we find a number of index choices for which terms of order higher than 
two occur, some of which coincide with the ones for odd $N$. As opposed to in the odd $N$ case, 
though, the number of possible choices for the summation indices depend on the parity of $i$, 
as well as on the modulo four value of $N$, so that the calculational details differ beteween, 
for instance, $N=6$ and $N=8$. This is because the restrictions one gets when solving the 
equal-indices-equations, are sometimes equivalent for, say, $i=3$ and $N=0$ mod $4$ and 
independent for $N=2$ mod $4$ for the same $i$, giving a different number of combinatorial 
possibilities. The final results for the frame potential average taken over Hilbert space, on 
the other hand, turns out to be identical for the two cases; this average is given in section 5.

Also when considering the sum average over ${\cal H}^+$ and ${\cal H}^-$, the calculations get more 
involved than in the non-restricted odd $N$ case. As for even $N$, we get cross term contributions 
to the integral, and further technical complications are due to the fact that the respective 
subspace conditions, 

\begin{equation} Z_i = Z_{-i} \end{equation}

\noindent and
\begin{equation} Z_i = -Z_{-i} \ , \end{equation}

\noindent make the number of equations to solve when asking what possibilities there are for the power 
of any $Z_i$ to be higher than two in a term larger. In other words, the calculations have to be 
divided into a greater number of different cases, but the logic is much the same as in the 
calculations for all of Hilbert space sketched above. In using equations (\ref{2-2-2-2} 
- \ref{8}), one should keep in mind that the dimension of the space over which the average is taken 
is no longer equal to $N$. 

\vspace{1cm}

{\bf 5. Results}

\vspace{5mm}

\noindent Our results are easy to state. The function $f$ attains its global 
minimum at a SIC, and its global maximum if all the $N^2$ unit vectors coincide. 
Thus 

\begin{equation} 0 \leq f \leq \frac{N^4}{2}\frac{N-1}{N+1} \ . \end{equation}

\noindent Its Fubini-Study average value is 

\begin{equation} \left< f\right>_{\rm FS} = \frac{N^2}{2}\frac{N-1}{N+1} 
\ . \label{fel} \end{equation}

\noindent When the 
dimension is large, the scalar product between randomly choosen vectors 
is close to zero, and this is reflected by the average. 
 
If we specialize to $N^2$ unit vectors forming an orbit under the 
Weyl-Heisenberg group we believe that the global maximum occurs if the 
fiducial vector is an eigenvector of some element in the group, in which case 
the orbit collapses to an eigenbasis. Thus 

\begin{eqnarray} 0 \leq f_H \leq \frac{N^3}{2}\frac{N-1}{N+1} \ . \nonumber 
\end{eqnarray}

\noindent (Unfortunately we were unable to prove that the eigenbasis represents 
the global maximum. It seems obvious that this is so, and we did check that it 
is a local extremum. For safety, we do not number this equation.) The average 
value depends on whether the Hilbert space dimension is odd or even:

\begin{equation} N = 2n - 1 \ : \hspace{8mm} \langle f_H\rangle_{\rm FS} = 
\frac{N^2}{2}\frac{N(N-1)}{(N+2)(N+1)} \end{equation}

\begin{equation} N = 2n \ : \hspace{8mm} \langle f_H\rangle_{\rm FS} = 
\frac{N^2}{2}\frac{N^2}{(N+3)(N+1)} \ . \end{equation}

\noindent Asymptotically there is no difference. It is perhaps somewhat 
unexpected that the average of $f_H$ has the same asymptotic behaviour as 
the average of $f$.  

We also computed the average when the fiducial vector is restricted to lie 
in certain interesting subspaces defined 
by elements of the Clifford group (the normalizer of the Weyl-Heisenberg 
group). Elements of order $2$ occur in odd dimensions $N = 2n-1$. Their 
eigenspaces have dimension dim$[{\cal H}^-] = n-1$ and dim$[{\cal H}^+] = 
n$. As observed by Hughston \cite{Hughston}, when $N = 3$ ${\cal H}^-$ 
contains only one vector, and it is a fiducial vector for a SIC.  
However, $3$ is a very special dimension. The complementary 
subspace ${\cal H}^+$ contains a SIC fiducial vector too, as well 
as four vectors from a complete set of four mutually unbiased 
bases, for which $f_H$ attains its (conjectured) maximum. The average over 
${\cal H}^+$ is $\left< f_H\right> = 81/40$. For $N > 3$ the situation is 
quite different. We find 

\begin{equation} N = 2n-1 > 3 \ : \hspace{8mm} 
\langle f_H\rangle_{{\cal H}^-} = \langle f_H\rangle_{{\cal H}^+} 
= N^2\frac{N(N-1)}{(N+3)(N+1)} \ . \end{equation}

\noindent Asymptotically this is twice the average over the full Hilbert space. 
It is a bit striking that the averages are equal. 
 
It is harder to deal with the subspaces defined by elements of order $3$. 
At the same time they are more interesting, because Zauner's conjecture 
\cite{Zauner, Marcus, Flammia} says that the fiducial vector can always be 
choosen to lie in the subspace with eigenvalue $1$. We did compute averages for $N = 7$. 
There are three subspaces ${\cal H}_{1}$, ${\cal H}_\alpha$, ${\cal H}_{\alpha^2}$, 
labelled by the eigenvalues of the element that cubes to one ($\alpha = e^{2\pi i/3}$). 
Their dimensions are respectively 3, 2, and 2. The average values of $f_H$ in the 
three subspaces are 

\begin{equation} \langle f_H\rangle_{{\cal H}_1} = \frac{151\cdot 7^3}{5\cdot 3^4\cdot 
2^3} \approx 15.985 \end{equation}

\begin{equation} \langle f_H\rangle_{{\cal H}_\alpha} = 
\langle f_H\rangle_{{\cal H}_{\alpha^2}} = \frac{37\cdot 7^3}{5\cdot 3^3\cdot 2^3} 
\approx 11.751 \ . \end{equation}

\noindent The average over the subspace that contains the SIC fiducial vector 
is above the average $\langle f_H\rangle$ over the full Hilbert space. This surprised 
us. 

The values of $f_H$ vary widely, also within the special subspaces (except 
that $f_H$ is actually constant on ${\cal H}^-$ when $N = 5$). We describe 
the situation for $N = 7$ in the following table:

\

\begin{center}

\begin{tabular}{|l||c|c|c|c|c|c|} \hline 
\  & $f$ & $f_H$ & $f_H({\cal H}^+)$ & $f_H({\cal H}^-)$ & $f_H({\cal H}_1)$ & 
$f_H({\cal H}_\alpha )$ \\ \hline
Minimum & 0 & 0 & 12.2 (?) & 4.764 (?) & 0 & ? \\
Average & 18.375 & 14.29 & 25.72 & 25.72 & 15.98 & 11.75 \\
Maximum & 900.4 & 128.6 (?) & 128.6 (?) & 42.88 (?) & ? & ? \\ 
\hline \end{tabular}

\end{center}

\

\

\noindent Some entries were left uncomputed, and some without proofs---we marked 
the latter with a question mark within bracket, even though we are sure they are correct. 

Unfortunately we are unable to see how to compute the asymptotic behaviour of 
$\langle f_H\rangle$ for a fiducial vector in subspaces of the kind considered by 
Zauner. We expect them to depend on number theoretical details of the dimension. 
It would be interesting to see numerical studies of such averages however.


\end{document}